\documentclass[conference]{IEEEtran}

\usepackage{cite}
\usepackage{amsmath,amssymb,amsfonts}
\usepackage{amssymb}
\usepackage{algpseudocode}
\usepackage{amsfonts}
\usepackage{graphicx}
\usepackage{fancyhdr}
\usepackage{cases}
\usepackage{textcomp}
\usepackage{extarrows}
\usepackage{algorithm,algpseudocode}
\usepackage{multirow}
\usepackage{graphicx}
\usepackage{overpic}
\usepackage{graphicx}
\usepackage{booktabs}
\usepackage{url}
\usepackage[caption=false]{subfig}
\IEEEoverridecommandlockouts

\usepackage{orcidlink}
\hypersetup{hidelinks}
\usepackage{multirow,tabularx}
\usepackage{mathtools}
\usepackage{xcolor}
\usepackage{listings}
\usepackage[english]{babel}
\usepackage{amsthm}
\usepackage{tcolorbox}
\tcbuselibrary{breakable}

\IEEEoverridecommandlockouts
\def\BibTeX{{\rm B\kern-.05em{\sc i\kern-.025em b}\kern-.08em
    T\kern-.1667em\lower.7ex\hbox{E}\kern-.125emX}}

\definecolor{customblue}{rgb}{0.0, 0.37, 0.63}

\begin{document}

\bstctlcite{BSTcontrol}
\title{TeleMoM: Consensus-Driven Telecom Intelligence via Mixture of Models}

\author{
    \IEEEauthorblockN{Xinquan Wang$^{1}$, Fenghao Zhu$^{1}$, Chongwen Huang$^{1}$, Zhaohui Yang$^{1}$, Zhaoyang Zhang$^{1}$,\\
    Sami Muhaidat$^{2}$, Chau Yuen$^{3}$,~\IEEEmembership{Fellow,~IEEE} and M\'{e}rouane~Debbah$^{4,}$$^{5}$,~\IEEEmembership{Fellow,~IEEE}
    }
    \IEEEauthorblockA{$^1$ College of Information Science and Electronic Engineering, Zhejiang University, 310027, Hangzhou, China}
    \IEEEauthorblockA{$^2$ Computer and Communication Engineering, Khalifa University, P.O. Box: 127788, Abu Dhabi, UAE}
    \IEEEauthorblockA{$^3$ School of Electrical and Electronics Engineering, Nanyang Technological University, Singapore 639798 }
    \IEEEauthorblockA{$^4$ KU 6G Research Center, Khalifa University of Science and Technology, P.O. Box 127788, Abu Dhabi, UAE}
    \IEEEauthorblockA{$^5$ CentraleSupelec, University Paris-Saclay, 91192 Gif-sur-Yvette, France}
    }

\maketitle
\begin{abstract}
Large language models (LLMs) face significant challenges in specialized domains like telecommunication (Telecom) due to technical complexity, specialized terminology, and rapidly evolving knowledge. Traditional methods, such as scaling model parameters or retraining on domain-specific corpora, are computationally expensive and yield diminishing returns, while existing approaches like retrieval-augmented generation, mixture of experts, and fine-tuning struggle with accuracy, efficiency, and coordination. To address this issue, we propose Telecom mixture of models (TeleMoM), a consensus-driven ensemble framework that integrates multiple LLMs for enhanced decision-making in Telecom. TeleMoM employs a two-stage process: proponent models generate justified responses, and an adjudicator finalizes decisions, supported by a quality-checking mechanism. This approach leverages strengths of diverse models to improve accuracy, reduce biases, and handle domain-specific complexities effectively. Evaluation results demonstrate that TeleMoM achieves a 9.7\% increase in answer accuracy, highlighting its effectiveness in Telecom applications.
\end{abstract}


\section{Introduction}\label{sec:intro}
The rapid evolution of large language models (LLMs) has revolutionized natural language processing, enabling unprecedented capabilities in reasoning, knowledge synthesis, and problem-solving. However, specialized domain like telecommunications (Telecom) domain presents unique challenges for LLMs due to its highly technical content, specialized terminology, and rapidly evolving knowledge base \cite{zhouhaoLLMTelecom, zhu2025wirelesslargeaimodel}. Yet, even state-of-the-art models exhibit limitations in handling Telecom knowledge base, where expertise demands precise integration of technical standards, academic research, and practical implementation \cite{shahid2025largescaleaitelecomcharting}. Traditional approaches, which rely on scaling model parameters or retraining on domain-specific corpora, face prohibitive computational costs and diminishing returns. Moreover, running state-of-the-art LLMs locally requires expensive high-end computing workstations, while reliance on model-as-a service providers raises concerns about data security and service reliability.
\par
These limitations underscore the urgent need for resource-efficient alternatives. Recent studies have explored several approaches to address this challenge, such as retrieval-augmented generation (RAG) \cite{RAG-KDD}, mixture of experts (MoE) \cite{MoETelecom} and fine-tuning \cite{zhang2024instructiontuninglargelanguage}. 
RAG enhances LLM performance by incorporating an external retrieval mechanism that fetches relevant documents or structured data before generating a response \cite{RAG-KDD}. Since RAG does not require full-scale model retraining, it provides an efficient way to enhance LLM responses with domain expertise. However, naive RAG pipelines may fall short in managing complex technical standards due to limitations in retrieval accuracy and contextual integration \cite{Telecom-RAG}. 
MoE is another promising approach by dynamically selecting a subset of neural networks for each query, thus reducing the overall processing load while improving domain-specific expertise \cite{MoETelecom}. However, MoE architectures pose challenges in model coordination, expert selection, and training complexity, particularly in Telecom applications where accurate handling of Telecom-specific queries depends on precise model routing and domain awareness.
Fine-tuning involves training a existing LLM on domain-specific datasets to enhance its expertise in a particular field \cite{zhang2024instructiontuninglargelanguage}. 
However, this approach entails significant computational costs, making it financially prohibitive. Furthermore, fine-tuning is highly contingent upon the availability of large, high-quality, domain-specific datasets. There is also the potential risk of overfitting, wherein the model becomes excessively specialized in the fine-tuning data, leading to a degradation of its general reasoning abilities and a diminished capacity to perform in scenarios not encompassed by the fine-tuning dataset. The lack of interpretability and trustworthiness is also a crucial challenge for real-world application \cite{robustnetwork, gmml}.
Currently, there are no effective solutions that adequately address these challenges.
\par
At the same time, different LLMs inherently possess unique strengths on different section of specialized domain like Telecom \cite{WEI2023}. 
For example, some may excel in protocol standard understanding, while others in signal processing theory or network optimization \cite{E220135,glnn,zhu2025liquidneuralnetworksnextgeneration}.
However, their fragmented expertise remains underutilized in isolation. This suggests an untapped opportunity by utilizing the intelligence and expertise of a mixture of models (MoM) by strategically blending their output through a framework that transcends individual capabilities. 
\par
To address this gap, we propose Telecom MoM (TeleMoM), a consensus-driven ensemble framework that integrates diverse LLMs for Telecom decision-making. TeleMoM operates in two stages: proponents generate responses with justifications, while an adjudicator evaluates and finalizes the decision. Additionally, we introduce a quality-checking and self-evaluation mechanism to enhance output reliability.
By leveraging multiple models, TeleMoM enhances accuracy, mitigates biases, and effectively handles domain-specific complexities. This structured approach improves interpretability and trustworthiness. In evaluations, TeleMoM increases answer accuracy score by 9.7\%, demonstrating its effectiveness.
\par
The remainder of this paper is structured as follows: Section \ref{sec:sys} presents the problem formulation, then Section \ref{sec:telemom} introduces the design of TeleMoM. Section \ref{sec:simulation} presents experimental results demonstrating the effectiveness of TeleMoM. Finally, section \ref{sec:conclusion} concludes this paper.

\section{Problem Formulation}\label{sec:sys}
The Telecom domain presents unique challenges for existing single general-propose LLM approaches due to its highly technical and expertise content \cite{ObservationsLLM}. Current strategies for domain adaptation each have drawbacks when applied to Telecom tasks. These methods usually require the use of single large general-purpose model to achieve high performance, leading to issues such as increased computational overhead. Moreover, both approaches still rely on internal reasoning of a single model, making them susceptible to hallucinations and limited by the inherent biases.
\par
Given these limitations, there is a clear need for a more robust solution. Instead of relying on a single model pipeline, we formulate the problem as an ensemble task. The research problem can be stated as: How can we design an ensemble of multiple LLMs that collaboratively yields accurate and robust solutions for Telecom tasks, surpassing the performance and domain coverage of any individual model? In other words, we seek an ensemble-based solution that mitigates single-model weaknesses such as limited knowledge or risk of errors by combining the strengths of diverse models.
\par
Formally, consider a question query $x$ in the Telecom domain (e.g., a question about network configurations, a user issue description, or a technical standard lookup). We employ a set of $N$ LLMs $\{M_1,M_2,\cdots, M_N\}$ as proponents. Each proponent can produce a candidate response $( y_i, r_i)=M_i(x)$, where $y_i$ is the proposed answer, and $r_i$ is the reason of the answer proposed that the proponent $i$ give. Let the ground truth answer to the query be $y^*$.
Our objective is to coordinate these multiple proposals in such a way that the final output $y$ approaches $y^*$ as closely as possible across a wide range of Telecom-related queries.
\par
To achieve this, we introduce an adjudicator $A(\cdot)$ which is itself an LLM. After receiving the query $x$ and all the proponent outputs $\{(y_1,r_1),(y_2,r_2),\cdots,(y_N,r_N)\}$, the adjudicator $A$ produces the final answer, denoted as
\begin{equation}\label{final_answer}
    y=A(x,(y_1,r_1),(y_2,r_2),\cdots,(y_N,r_N)).
\end{equation}

Finally, we introduce a evaluation mechanism $S(y,y^*)$ that scores the final output $y$ based on its similarity with ground truth $y^*$, and the entire problem could be presented as
\begin{equation}
    \max S(y,y^*).
\end{equation}

This formulation sets the stage for a structured ensemble approach, where multiple LLMs collaboratively refine and validate answers to enhance accuracy, robustness, and domain coverage.
\section{Consensus-driven Mixture of Models for Telecom Knowledge}\label{sec:telemom}
In this section, we describe how TeleMoM is designed, how it manages the interactions among models, and why this ensemble approach significantly improves resilience, accuracy, and domain coverage in Telecom applications.

\subsection{Mixture of Models Structure in TeleMoM}
TeleMoM is based on the MoM paradigm, where multiple models with different specializations collaborate to provide robust and accurate answers.
The key advantage of MoM in TeleMoM is its ability to systematically integrate knowledge from various knowledge bases and sources across the models. By structuring interactions between different models, TeleMoM mitigates single-model failure points and ensures that decisions are made based on a broader knowledge spectrum.
\par
The proposed TeleMoM framework consists of two primary components, an advisory committee and an adjudicator model, which are arranged in an ensemble architecture presented in Fig. \ref{fig:telemom}, and example prompts for the proponents and the adjudicator can be found in Appendix \ref{appendix1}.
\begin{itemize}
    \item The advisory committee consists of multiple LLMs, referred to as proponents, each attempt to solve the query potentially from their respective knowledge bases or domains of expertise.
    \item The adjudicator is also a LLM responsible for evaluating and synthesizing responses from the advisory committee. It determines the most accurate, reliable, and comprehensive final answer by either selecting the best response or integrating multiple responses. The adjudicator works without additional fine-tuning. 
\end{itemize}

\begin{figure}[t]
	\begin{center}
		\centerline{\includegraphics[width=\linewidth]{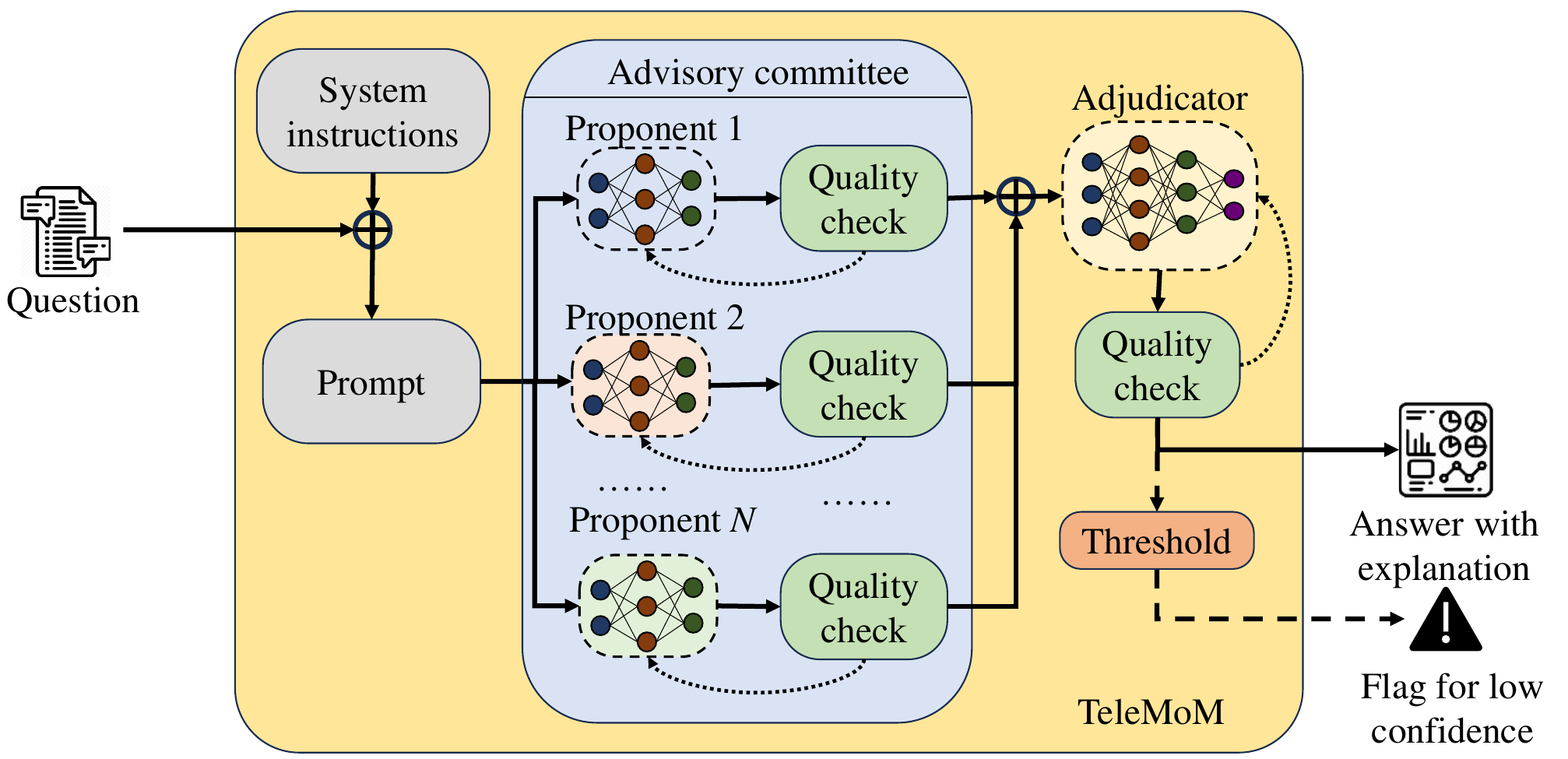}}  
		\captionsetup{font=footnotesize, name={Fig.}, labelsep=period}
		\caption{The TeleMoM architecture.}
		\label{fig:telemom}
	\end{center}
	\vspace{-12mm}
\end{figure}

\subsection{Proponent--Adjudicator Interaction}\label{sec:inter}
\subsubsection{Parallel Query to Proposers}
When a Telecom-related query $x$ is received, TeleMoM initiates its workflow by distributing $x$ to all $N$ proponents in the advisory committee in parallel (or in sequence when the computing resource is limited, since all proponents work independently). Each proponent $M_i$, which may be a general LLM, a Telecom-specialized LLM, or an LLM with RAG, receives specific instructions to produce high-quality response for the adjudicator from a instructor. Each proponent then generates an output pair $(y_i, r_i) = M_i(x)$, where $y_i$ represents the candidate response and $r_i$ provides supporting rationale or references. An automatic check is performed after the proponents provide their responses, evaluating the quality of the response (eg. format, length, completeness, etc.). If a response does not meet predefined criteria, the proponent is required to redraft and resubmit its answer before proceeding to the next stage.
As an optional feature, we introduce a confidence leveling mechanism from the inspiration of \cite{stangel2025rewarding}. Each proponent can also provide a score indicating its certainty in the response, which is later categorized into high, medium, or low level based on a threshold.
This iterative mechanism ensures that each candidate response adheres to high-quality standards before being forwarded to the adjudicator.

\subsubsection{Adjudication of Candidate Outputs}
All candidate outputs ${(y_1, r_1), (y_2, r_2), \dots, (y_N, r_N)}$ are sent along with the original query $x$ to the adjudicator. The adjudicator evaluates these responses by cross-referencing and comparing them.
If a consensus emerges among the proponents, i.e. multiple proponents provide highly similar responses, the adjudicator selects the best single answer. Alternatively, if the responses vary significantly, the adjudicator synthesizes a composite response by holistically considering the reasoning provided by the advisory committee and making its own judgment. As an optional feature, if major discrepancies exist, or weak consensus among proponents exist, the adjudicator may flag uncertainty by giving low or medium confidence levels following a similar mechanism in proponents.

\subsubsection{Decision and Final Output}
Upon completing the adjudication process, the adjudicator produces the final response $y$. Depending on the nature of the query, the adjudicator may also append relevant references, explanations, or justifications to support its decision. Depending on the nature of the query, the adjudicator may also append relevant references, explanations from the advisor committee, or justifications to support its decision.
The final answer is then presented to the user in a structured and interpretable format. In cases where confidence levels are low or ambiguity remains, the adjudicator may flag the response for further human verification or provide alternative perspectives for the consideration of the user.

\begin{figure}[t]
    \centering
	\includegraphics[width=\linewidth]{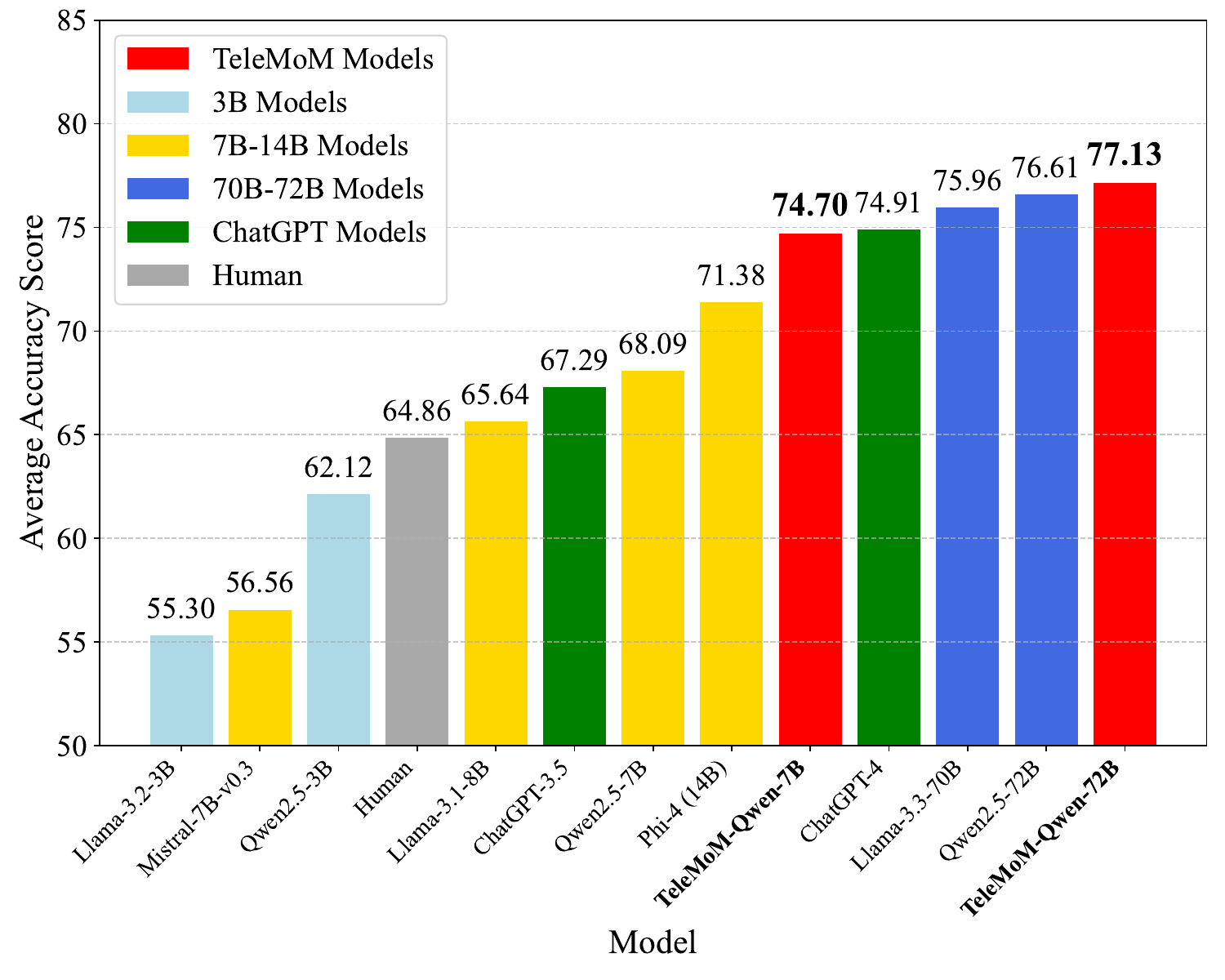}\vspace{-4mm}
	\captionsetup{font=footnotesize, name={Fig.}, labelsep=period}
    \caption{TeleMoM vs. baseline models on accuracy score.}
    \vspace{-4mm}
    \label{fig:accuracy}
\end{figure}

\begin{figure}[t]
    \centering
	\includegraphics[width=\linewidth]{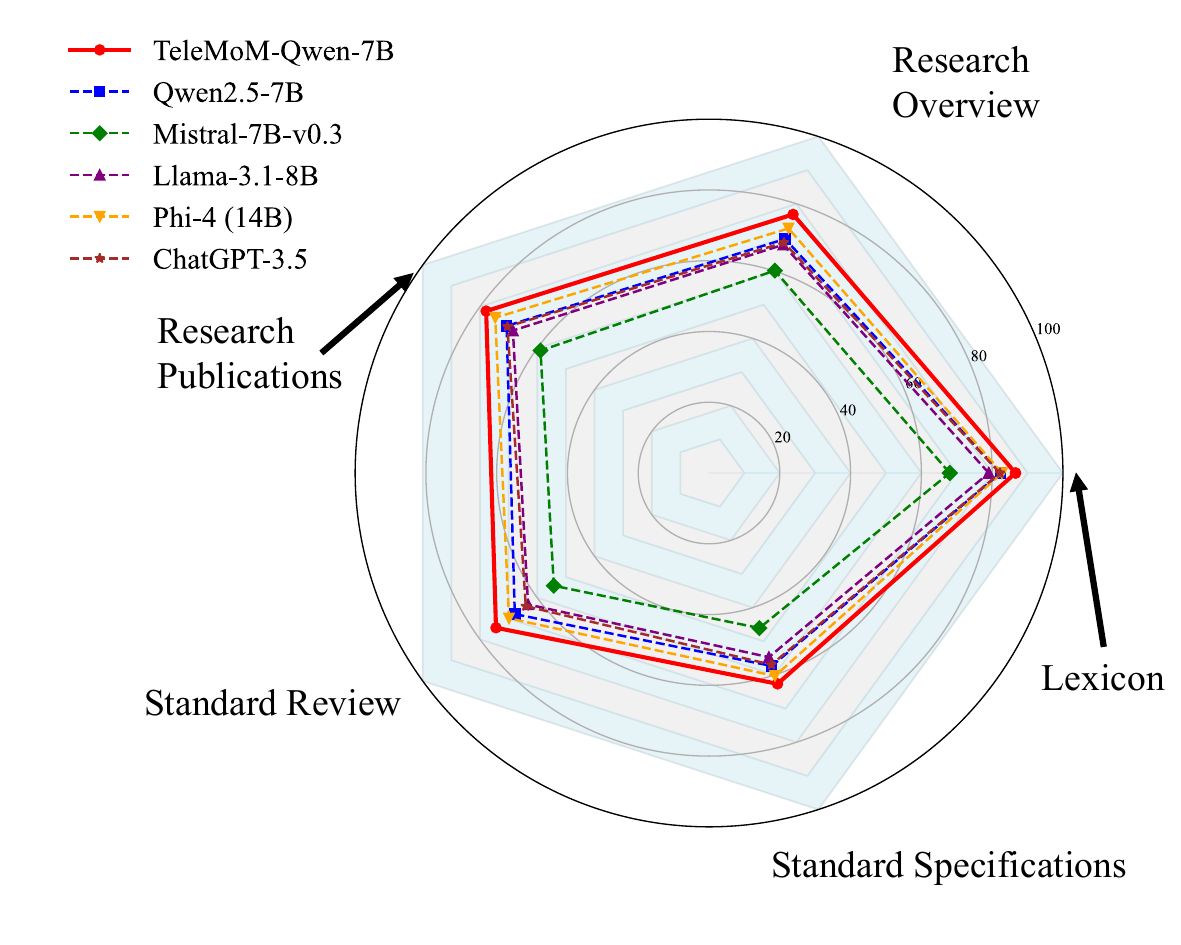}\vspace{-4mm}
	\captionsetup{font=footnotesize, name={Fig.}, labelsep=period}
    \caption{Average accuracy score comparison of models in all five categories.}\vspace{-6mm}
    \label{fig:category}
\end{figure}

\section{Evaluation Results}\label{sec:simulation}
In this section, we provide a comprehensive evaluation of the proposed TeleMoM framework across various Telecom-related topics. Evaluation is performed using the TeleQnA dataset \cite{maatouk2023teleqna}, a corpus of 10,000 Telecom questions structured into five major categories, detailed below:
\begin{itemize}
    \item \textit{Standards Specifications}: Detailed documents defining technical standards for Telecom systems, sampled uniformly across standardization bodies.
    \item \textit{Standards Overview}: Broader materials on summaries, review publications, and white papers that offer a broad perspective on Telecom standards beyond technical specifications.
    \item \textit{Research Publications}: Technical insights derived from research articles and open-access technical books, with questions that require LLMs to demonstrate an in-depth understanding of Telecom topics.
    \item \textit{Research Overview}: Content derived from review and survey publications that analyze and summarize research findings in the field.
    \item \textit{Lexicon}: A collection of technical terms extracted from standards and research documents to assess the understanding of Telecom-specific terminology of the models.
\end{itemize}

In the evaluation, TeleMoM uses four open-source LLMs as proponents: Qwen2.5-7B \cite{qwen2025qwen25technicalreport},
Llama 3 \cite{grattafiori2024llama3herdmodels},
Mistral \cite{choi2024linqembedmistraltechnicalreport},
and Phi-4 \cite{abdin2024phi4technicalreport}. Evaluation results of ChatGPT and Human experts were drawn from \cite{maatouk2023teleqna}.
For the evaluation, we use Python 3.10 and SGLang \cite{SGLang} as inference framework.
For the adjudicator, we use Qwen2.5 series and use Qwen2.5-7B as the default model. For clarity, we denote TeleMoM with Qwen2.5-7B and Qwen2.5-72B adjudicator as TeleMoM-Qwen-7B and TeleMoM-Qwen-72B, respectively.

\begin{figure}[t]
    \centering
	\includegraphics[width=\linewidth]{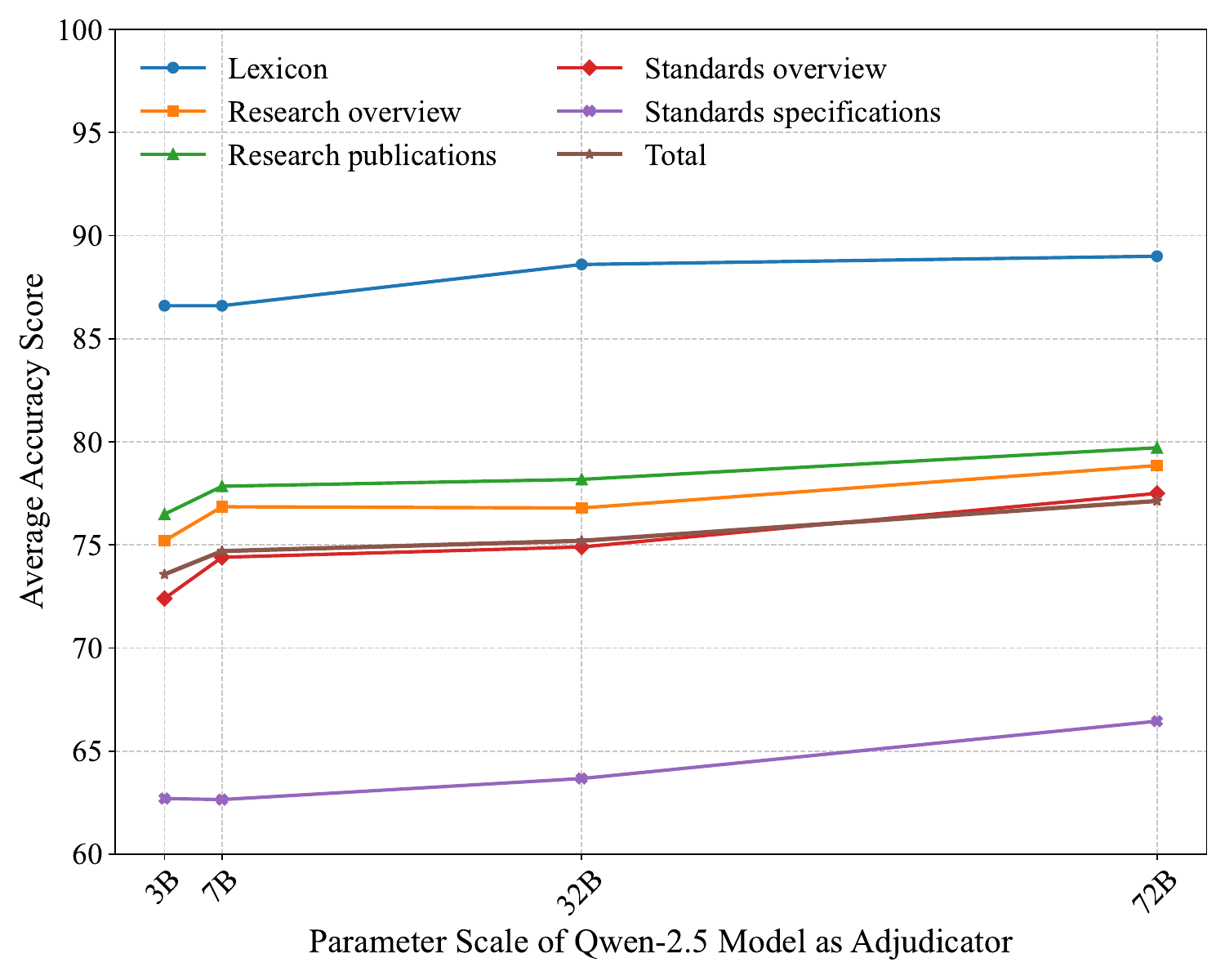} \vspace{-8mm}
	\captionsetup{font=footnotesize, name={Fig.}, labelsep=period}
    \caption{Average accuracy score vs. adjudicator model size.}\vspace{-4mm}
    \label{fig:adjudicator_scaling}
\end{figure}

\begin{figure}[t]
    \centering
	\includegraphics[width=\linewidth]{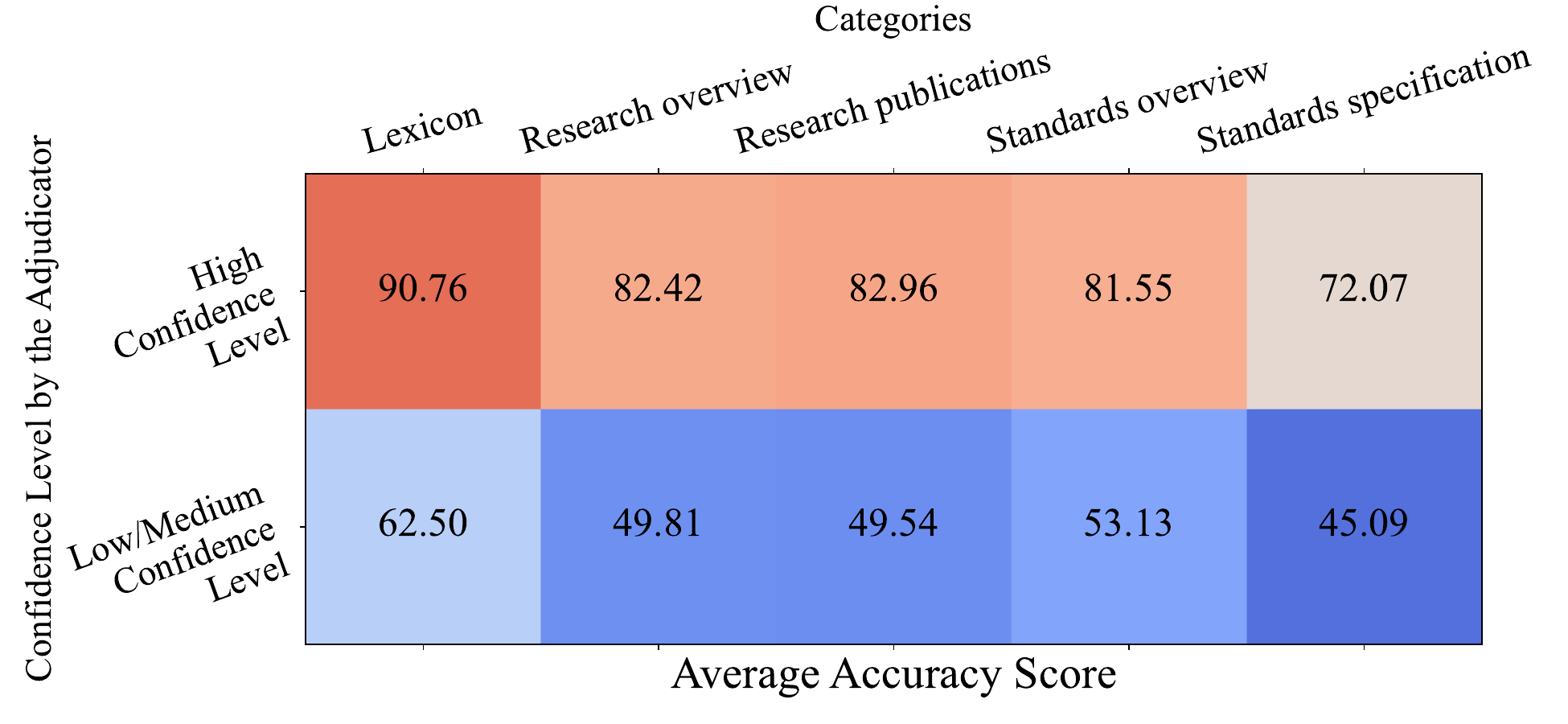}\vspace{-4mm}
	\captionsetup{font=footnotesize, name={Fig.}, labelsep=period}
    \caption{Average accuracy on different confidence level by the adjudicator. The number in the figure denotes the average accuracy score in the respective confidence level by the adjudicator and category.}
    \vspace{-4mm}
    \label{fig:confident}
\end{figure}

\subsection{Comparison with Individual Models}\label{sec:sim1}
To evaluate the effectiveness of TeleMoM, we compare its performance against individual baseline models. The results are illustrated in Fig. \ref{fig:accuracy}. It is demonstrated that TeleMoM-Qwen-7B and TeleMoM-Qwen-72B, which incorporates Qwen-7B and Qwen-72B as adjudicators respectively, achieve significant performance improvements. Specifically, TeleMoM-Qwen-7B attains a average accuracy score of 74.70, markedly surpassing Qwen-7B itself of 68.09, which is 9.7 \% higher. Its performance is also comparable to Llama-3.3-70B, which scores 75.96. Meanwhile, TeleMoM-Qwen-72B achieves an even higher score of 77.13, outperforming Qwen2.5-72B itself.
This substantial improvement underscores the benefits of integrating multiple specialized LLMs within the TeleMoM framework. By leveraging the strengths and sharing the knowledge base of different models, TeleMoM enhances response accuracy beyond what individual models can achieve independently.
\par
Next, we analyze the performance of TeleMoM across the five TeleQnA categories, as depicted in Fig. \ref{fig:category}. The results indicate that TeleMoM-Qwen-7B outperforms all compared models in every category, even surpassing Phi-4 (14B parameters). Notably, all models achieve their highest performance in the lexicon category and their lowest in standard specifications. This trend may stem from the fact that standard specifications require a high degree of specialized expertise and are less connected to common knowledge. The performance gains observed with TeleMoM confirm its ability to effectively consolidate insights from diverse LLMs, outperforming individual models while maintaining efficient inference performance. Its multi-model collaboration approach proves particularly advantageous for tackling complex Telecom queries that demand both high precision and deep contextual understanding. For a detailed comparison of answer accuracy score across various models and categories, please refer to Table \ref{detailed_scores} in Appendix \ref{appendix2}.

\subsection{Evaluation with Different Size of Adjudicator}
We further investigate the impact of adjudicator model size on the performance of TeleMoM by experimenting with different model scales. With the default settings, we select the Qwen-2.5 series spanning from Qwen2.5-3B to Qwen2.5-72B ad adjudicators to analyze the relationship between adjudicator size and the overall effectiveness of TeleMoM. The results presented in Fig. \ref{fig:adjudicator_scaling} reveal varying levels of improvement across different question categories.
Among all categories, standard overview exhibits the highest performance gain with an increase of 5.1 points, while research overview improves by 3.65 points. This trend suggests that as the adjudicator scales up, it becomes more proficient at analyzing and synthesizing information from the advisory committee, leading to enhanced decision-making accuracy.
The observed gains across different categories suggest that TeleMoM benefits from a hierarchical reasoning process in which the adjudicator carefully weighs diverse perspectives and resolves inconsistencies among proponents to reach well-informed decisions.

\par
\subsection{Evaluation Incorporating Confidence Levels}
As an optional extension discussed in Section \ref{sec:inter}, TeleMoM can incorporate confidence levels from proponents to inform the adjudicator about their certainty in their responses. Additionally, the adjudicator can flag responses as medium or low confidence level when it detects significant disagreements among proponents or notable discrepancies.
\par
To evaluate this mechanism, we conducted an experiment as follows. The adjudicator now receives confidence level from the proponents and requested to flag its confidence on its output.
In this experiment, we employ Qwen2.5-72B as the adjudicator.  
The results, illustrated in Fig. \ref{fig:confident}, show a significant improvement in high-confidence categories. For instance, in the standard specification category, responses with high confidence achieved an average accuracy score of 72.07, a substantial increase from the original accuracy score of 66.45. This demonstrates that the confidence-aware adjudication mechanism effectively identifies cases that may require external verification or human review, enhancing the reliability of the TeleMoM.

\section{Conclusion}\label{sec:conclusion}
In this paper, the TeleMoM framework was introduced to tackle the challenges of applying LLMs to the Telecom domain. A consensus-driven ensemble of multiple LLMs was employed, enhancing accuracy and reliability in addressing Telecom-related queries. Through evaluations, a 9.7\% increase in accuracy score was achieved, demonstrating the framework’s effectiveness. The structured approach, integrating proponent models and an adjudicator, was shown to strengthen interpretability and trustworthiness, marking a valuable advancement in LLM-driven Telecom knowledge applications.

\begin{appendices}

\section{Example Prompts for the Proponents and the Adjudicator}\label{appendix1}
The prompts for the proponents and adjudicator in TeleMoM are designed to guide their specific roles. We provide example prompts for the proponents and adjudicator in TeleMoM in this appendix. 

First, the prompt for the proponents are given. As domain experts, proponents are required to offer structured and detailed answers in the required format. This promotes consistency and clarity, allowing for efficient quality checks and enabling the adjudicator to easily process and evaluate the responses.

\begin{tcolorbox}[colback=blue!10!white,colframe=customblue,title=Example Prompt for the Proponents]
You are an expert in an telecommunication technical committee. Your role is to give suggestion to the adjudicator who make final decisions.\\
Please provide the answers to the following telecommunications related questions. The questions will be in a JSON format, the answers must also be in a JSON format as follows:\\
{[}format requirement{]}\\
Question: [question description] 
\end{tcolorbox}

Next, we present the prompt for the adjudicator, which emphasizes the need to analyze and synthesize responses from multiple proponents. This structured approach helps the adjudicator can make informed and accurate decisions based on the provided responses and justifications.

\begin{tcolorbox}[colback=blue!10!white,colframe=customblue,title=Example Prompt for the Adjudicator]
You are an expert in telecommunication field, good at analyzing and giving answers to complicated questions. \\
Based on the information given below, answers the question in the telecommunication field.\\
Question: [question description] \\
Answer by each model and reason: \\
{[}model 1 name{]}: {[}answer by model 1{]}\\
{[}reason 1{]}: {[}reason by model 1{]}\\
$\cdots \cdots$\\
Analyse the information given, and give your answer.\\
Respond in JSON with the following structure: \\
{[}format requirement{]}
\end{tcolorbox}

These examples serve as templates to demonstrate the expected format and logic. In practice, prompt wording can be adapted based on task complexity, domain specificity, or model behavior. They are not intended as fixed designs but as starting points for building more refined prompting strategies.

\section{Detailed Benchmarking Results}\label{appendix2}
We provide a detailed accuracy comparison for various models across different categories discussed in Section \ref{sec:sim1}, as shown in Table \ref{detailed_scores}. For clarity, here we use Le, RO, RP, SO, and SS to denote lexicon, research overview, research publications, standard overview, and standard specifications, respectively.

According to the result, model size has a notable impact on performance, particularly in Le and SO. Larger models like Llama-3.3-70B and Qwen2.5-72B outperform smaller ones in these areas, indicating that larger models better grasp specialized terminology and complex standards. This trend continues with RP and RO where the larger models excel. Smaller models tend to underperform in these categories, struggling to process specialized knowledge. In SS, which requires deep understanding of Telecom standards, large models such as TeleMoM-Qwen-72B excel. Smaller models like Llama-3.2-3B fall behind, likely due to their reduced capacity to handle technical intricacies. However, TeleMoM-Qwen-7B stands out with an average accuracy improvement of 9.7\% over its baseline counterparts, showcasing the advantage of using multiple models to synthesize responses and improve overall accuracy.
The results highlight that combining multiple models within the TeleMoM framework leads to improved performance in Telecom-related tasks.

\begin{table}[t]
    \centering
    \caption{Models Average Accuracy Score Across TeleQnA Categories}
    \renewcommand{\arraystretch}{1.3}
    \begin{tabular}{@{}l cccccc@{}}
        \toprule
        Model Name & Le & RO & RP & SO & SS & Avg. \\
               \midrule
        Llama-3.2-3B & 67.94 & 59.29 & 57.77 & 52.38 & 45.51 & 55.30 \\
        Llama-3.1-8B & 79.05 & 67.83 & 68.50 & 63.20 & 54.72 & 65.64 \\
        Llama-3.3-70B & 88.60 & 78.70 & 78.33 & 76.50 & 64.43 & 75.96 \\
        Qwen2.5-3B & 79.79 & 65.37 & 64.18 & 61.26 & 50.99 & 62.12 \\
        Qwen2.5-7B & 82.36 & 69.55 & 70.71 & 67.84 & 57.34 & 68.09 \\
        Qwen2.5-72B & 88.80 & 77.45 & 78.88 & 76.82 & 67.48 & 76.61 \\
        Mistral-7B-v0.3 & 68.00 & 60.15 & 58.87 & 54.20 & 46.07 & 56.56 \\
        Phi-4 & 82.77 & 72.68 & 74.67 & 69.86 & 60.30 & 71.38 \\
        ChatGPT-3.5 & 82.20 & 68.50 & 70.42 & 64.00 & 56.97 & 67.29 \\
        ChatGPT-4 & 86.80 & 76.25 & 77.62 & 74.40 & 64.78 & 74.91 \\
        Human & 80.33 & 63.66 & 68.33 & 61.66 & 56.33 & 64.86 \\
        \midrule
        \textbf{TeleMoM-Qwen-7B} & 86.60 & 76.85 & 77.84 & 74.40 & 62.65 & 74.70 \\
        \textbf{TeleMoM-Qwen-72B} & 89.00 & 78.85 & 79.71 & 77.50 & 66.45 & 77.13 \\
        \bottomrule
    \end{tabular}\vspace{-4mm}
    \label{detailed_scores}
\end{table}

\section{Visualization of an Adjudication Example}
To illustrate the TeleMoM workflow, we present a concrete example. In this scenario, the TeleMoM models are given a multiple choice question, and their suggested answers, along with the reasoning behind each selection, are displayed in the table. 
Among the models, all except Mistral-v0.3-7B recommended selecting option 3, while Mistral-v0.3-7B chose option 4. Ultimately, the adjudicator aligned with the consensus of the other three models.

\begin{tcolorbox}[colback=blue!10!white,colframe=customblue,title=Example Adjudication Process,breakable]
\textbf{Question}: \\
What do multiple cooperative half-duplex (HD) eavesdroppers do in UAV (Unmanned aerial vehicle) wireless communication systems? Option 1: They transmit jamming noise to the legitimate receiver, option 2: They intercept the confidential signals only, option 3: They transmit jamming noise and intercept the confidential signals simultaneously, option 4: They mimic a full-duplex eavesdropper without transmitting jamming.\\
\textbf{Qwen-2.5-7B}: [Selected option 3] \\
Reason: In a cooperative half-duplex eavesdropping scenario, the eavesdroppers not only intercept the confidential signals for potential malicious use or intelligence gathering, but they also transmit jamming noise to interfere with the legitimate communication, potentially harming the intended receiver's reception of the signal.\\
\textbf{Phi-4} (14B): [Selected option 3] \\
Reason: Cooperative half-duplex eavesdroppers in UAV communication systems often perform both jamming and interception to disrupt the legitimate receiver while capturing the confidential signals.\\
\textbf{Llama-3.1-8B}: [Selected option 3] \\
Reason: This is the correct option as multiple cooperative half-duplex eavesdroppers are known to simultaneously transmit jamming noise and intercept confidential signals in UAV wireless communication systems.\\
\textbf{Mistral-v0.3-7B}: [Selected option 4] \\
Reason: Cooperative HD eavesdroppers in UAV communication systems can both transmit jamming signals to make it difficult for the legitimate receiver to receive the signal and intercept the confidential signals to gather information. \\
\textbf{Adjudicator} (Qwen-2.5-7B): [Selected option 3] \\
Reason: The majority of the models consistently select option 3, indicating that multiple cooperative half-duplex eavesdroppers in UAV wireless communication systems both intercept confidential signals and transmit jamming noise simultaneously. This dual-action approach disrupts legitimate communication while allowing eavesdroppers to capture sensitive information. The consensus among these models, alongside the detailed reasoning provided, strongly suggests that option 3 is the most accurate choice.

\end{tcolorbox}

\end{appendices}

\bibliographystyle{IEEEtran}
\bibliography{TeleMoM}

\end{document}